\documentstyle[epsf, twoside]{esapub}

\newcommand{\gw}{gravitational wave}
\newcommand{\grad}{gravitational radiation}
\newcommand{\dtc}{detector}

\newcommand{\msolar}{M_{\odot}}

\newcommand{\mptr}{{\cal M}}

\newcommand{\fracparen}[2]{\left(\frac{#1}{#2}\right)}
\newcommand{\fracsqb}[2]{\left\lbrack\frac{#1}{#2}\right\rbrack}

\newcommand{\ibar}{\mbox{\rlap{$I$}--}}

\newcommand{\oderiv}[2]{\frac{d #1}{d #2}}
\newcommand{\oderivn}[3]{\frac{d^{#3}\!#1}{d #2^{#3}}}

\newcommand{\E}[1]{\times 10^{#1}}

\newcommand{\splA}[2]{#2}
\newcommand{\splB}[2]{#1}
\newcommand{\American}{\let\spl=\splA} 
\newcommand{\British}{\let\spl=\splB} 

\newcommand{\Eqref}[1]{Equation~(\ref{#1})}
\newcommand{\Figref}[1]{Figure~\ref{#1}}

\newcommand{\hide}[1]{}\newcommand{\units}{\rm\;}

\newcommand{\Cardiff}{Department of Physics and Astronomy, University of Wales College of Cardiff, Cardiff, UK}
\newcommand{\AEI}{Max Planck Institute for Gravitational Physics (Albert Einstein Institute), Potsdam, Germany}

\begin{document}

\setlength{\parindent}{0pt}
\setlength{\parskip}{ 10pt plus 1pt minus 1pt}
\setlength{\hoffset}{-1.5truecm}
\setlength{\textwidth}{ 17.1truecm }
\setlength{\columnsep}{1truecm }
\setlength{\columnseprule}{0pt}
\setlength{\headheight}{12pt}
\setlength{\headsep}{20pt}
\pagestyle{esapubheadings}
\twocolumn
\title{\bf LOW-FREQUENCY SOURCES OF GRAVITATIONAL WAVES: A TUTORIAL\thanks{%
To be published in the Proceedings of the 1997 Alpbach Summer School on Fundamental 
Physics in Space, ed. A Wilson, ESA (1997).}}

\author{{\bf B.~F.~Schutz} \vspace{2mm} \\
{\bf \AEI} \and {\bf \Cardiff } }

\maketitle

\begin{abstract}

Gravitational wave \dtc s in space, particularly the LISA project, can study a rich variety of astronomical systems whose \grad\ is not detectable from the ground, because it is emitted in the low-frequency gravitational wave band (0.1~mHz to 1~Hz) that is inaccessible to ground-based detectors.  Sources include binary systems in our Galaxy and massive black holes in distant galaxies.  The radiation from many of these sources will be so strong that it will be possible to make remarkably detailed studies of the physics of the systems.   These studies will have importance both for astrophysics (most notably in binary evolution theory and models for active galaxies) and for fundamental physics.  In particular, it should be possible to make decisive measurements to confirm the existence of black holes and to test,  with accuracies better than 1\%, general relativity's description of them.   Other observations can have fundamental implications for cosmology and for physical theories of the unification of forces.  In order to understand these conclusions, one must know how to estimate the gravitational radiation produced by different sources.  In the first part of this lecture I review the dynamics of gravitational wave sources, and I derive simple formulas for estimating wave amplitudes and the reaction effects on sources of producing this radiation.   With these formulas one can estimate, usually to much better than an order of magnitude, the physics of most of the interesting low-frequency sources.  In the second part of the lecture I use these estimates to discuss, in the context of the expected sensitivity of LISA, what we can learn by from observations of binary systems, massive black holes, and the early Universe itself.  \vspace {5pt} \\


  Key~words: gravitational waves, LISA, black holes

\end{abstract}

\section{INTRODUCTION}

The LISA project, currently identified as a  future European Space Agency Cornerstone mission in the next century, is one of the most ambitious scientific observatory missions ever contemplated.   To detect gravitational waves one needs to build an instrument capable of measuring picometre-scale changes in the distances between spacecraft separated by millions of kilometres.  The fascinating and challenging technology to do this will be explored in other lectures at this school, and reviews can also be found in the literature (\cite{danzmann,LISA1996}).  The motivation to build LISA, or for that matter any other space-based detector of gravitational radiation, lies in its scientific return: what will LISA learn?  

Any astronomical observation is partly a ``lucky dip'', the game where one puts one's hand into a sack and pulls out an object at random.  One never knows for sure what an instrument will reveal about a part of the Universe that one has never looked at before.  This is especially true for the first observations in a new waveband, and it will surely be as true for gravitational wave detectors as it has been for radio antennas, X-ray satellites, and underground neutrino detectors.  Nevertheless, large projects like LISA cannot be justified on serendipity --- on our ignorance! --- alone.  We must try to estimate as carefully as possible the kinds of gravitational wave sources LISA might see and what information is likely to be deducible from their observation.

LISA will operate in what we call the low-frequency band of gravitational waves, between $1\E{-4}\units Hz$ and 1~Hz.  We will see that this is an interesting band, where we expect radiation from binary stars, massive black holes, and possibly the Big Bang.  The sources that radiate strongly in this band are very different from those that radiate in the higher-frequency band from about 1~Hz to $10^4\units Hz$, which Earth-based detectors have targeted since the early 1960's.  Most of them do not radiate at the higher frequencies, and so much of the physics and astrophysics that LISA can explore will be totally new, even if (as we hope) ground-based instruments are successful soon in making the first detections of gravtational waves.  

Importantly, the low-frequency band cannot be observed from the ground.  Any gravitational wave detector is a sensitive recorder of time-dependent changes in the gravitational field, and it will respond as well to changes in the local Newtonian gravitational field induced by the motion of a terrestrial mass (a person or a truck) as to a gravitational wave.  Because gravitational fields cannot be screened out, it is impossible to avoid terrestrial interference.  

In the higher-frequency band from about 1~Hz to $10^4\units Hz$, terrestrial gravitational noise is smaller than the signals from astronomical sources, and detectors can be built on the ground.  In the LISA band, the reverse is true: gravitational interference from local mass movements, from density perturbations carried by seismic activity, and even from the passage of atmospheric masses is much stronger than extraterrestrial waves, and Earth-based detection is hopeless.  By putting a detector in space, far enough from Earth, one escapes this interference: terrestrial noise falls off in strength moving away from Earth, while the amplitude of the incoming gravitational waves is essentially the same everywhere in the solar system, since they come from so far away.

In space, the natural method of detecting gravitational waves is interferometry.  Once one leaves the near-Earth environment to build such a detector, one can take advantage of a great bonus: the vacuum system is free.  Gravitational waves act by tidal forces, so the displacements they make in a detector are proportional to the size of the detector.  Ground-based interferometers are limited in size to a few kilometres simply by the cost of building a vacuum system of that size; ideally they should be several hundred kilometres in length.  In space one can make as large a detector as one likes, within technological constraints.   

The result is that detectors need not be designed to have only barely enough sensitivity to detect something: LISA will be so sensitive that its signal-to-noise ratio when it detects the collision of two massive black holes in a distant galaxy could well be better than that of an optical observation of the same galaxy.  It will have enough signal to pin down directions to sources, to measure their masses and other properties, and especially to look for the small details in the signal that will test our understanding of aspects of fundamental physics.  When thinking about LISA, one must forget the impression that one  has from ground-based gravitational wave projects, that gravitational wave detectors operate at the margins of detection.  LISA will be a robust observatory.  

Since this is a summer school, it is important that students learn more than what the experts think about LISA and its capabilities.  Students should be able to do their own source calculations, making their own assessment of the capabilities of the instrument and the interest of the physics in this wave band.  It is not in fact difficult to give rough (factor of two) estimates of the radiation amplitudes expected from sources and of the back-reaction on the sources of losing energy to gravitational radiation.  I will show in 
this lecture how to derive such estimates from the fundamentals of Newtonian gravity and general 
relativity.  These include some simple approximations, such as \Eqref{eqn:hbound} and 
\Eqref{eqn:luminosity}, that have been known for some time but deserve to be more widely used.  I assemble a number of these estimates into a single diagram, \Figref{fig:sourcedynamics}, that shows 
the dynamics of sources as a function of their mass and size: what their radiation frequency will be, 
which ones will be strongly affected by radiation-reaction effects within a one-year observation, and which ones will be affected less strongly but still measurably by such effects.  Most of this section is 
equally applicable to sources that radiate in the high-frequency band as well as to those detectable 
at lower frequencies.

After this introduction to the physics of sources, I will connect what we have learned about the physics 
of sources to the capabilities of the LISA instrument.  I will discuss what we can learn about binaries, 
massive black holes, cosmology, and fundamental physics by making observations in the LISA band.
The rich nature of the information we expect to receive provides the underpinning motivation for LISA.  
This lecture is the companion to my second lecture at this school, which reviews methods of data 
analysis for gravitational radiation and uses them to deduce the performance of LISA in deducing  
information about the sources discussed here.

Students wanting further discussions of ground- and space-based sources can turn to a number of 
references: \cite{thorne1987,thornesnowmass,leshouches}.

\section{PHYSICS OF GRAVITATIONAL WAVE SOURCES}\label{sec:dyn}

The gravitational wave spectrum of space- and ground-based detection spans 8 orders of magnitude in frequency, from $10^{-4}$~Hz to $10^4$~Hz.  This is similar to the range from high-frequency radio waves (10~GHz)  to X-rays ($10^{18}$~Hz).  In this range, therefore, we should expect considerable variety.  But there is also a lot that is systematic.  The dynamics of most sources are dominated by their self-gravity, and their gravitational-wave amplitudes will usually be given to a good approximation by the lowest-order quadrupole approximation for radiation.  

\subsection{Internal dynamics: the natural frequency}

For self-gravitating Newtonian systems, it is well-known that there is a natural dynamical frequency associated with the mean mass density $\rho$ of the system: 
\begin{equation}\label{eqn:naturalf}
f_{\rm dyn}=\frac{1}{2\pi}\left(\pi G\rho\right)^{1/2} \sim \fracparen{GM}{16\pi R^3}^{1/2},
\end{equation}  
where $M$ is the system's mass and $R$ its typical size. Wherever I use the symbol $\sim$ instead of $=$, I mean to indicate that there are factors of order 2 or pi left out.\footnote{I am grateful to Prof.\  W.\ Kummer for telling me at this meeting that such a factor of order one that is omitted from an expression is sometimes called a Weisskopfian, after Victor Weisskopf, who perfected this style of calculation!} (In this case, the factor is $\pi/3$.)  Moreover, I will always use proper frequencies (measured in Hz), not angular frequencies (radians/s) in my formulas.  

It is interesting to put some numbers into this formula.  Selecting values relevant to LISA's sources, it is 
easy to show that:
\begin{equation}\label{eqn:naturalfnumbers}
f_{\rm dyn} =1\E{-3} \fracsqb{M}{2.8\msolar}^{1/2}\fracsqb{R}{2\E{8}\units m}^{3/2}\quad\units Hz.
\end{equation}
Similarly, solving \Eqref{eqn:naturalf}  for the density, we get
\begin{equation}\label{eqn:rhofromfreq}
\rho \sim 2\E{6}\fracsqb{f}{3\units mHz}^2 \units kg\;m^{-3}.
\end{equation}
Notice that, for frequencies in the range $10^{-4}$ -- $10^4$~Hz, the density ranges from nuclear-matter density at the high end down to the density of water at the low end.  This illustrates the enormous range of physics in these sources.  

This formula for the relation between frequency and density is valid to a first approximation in general relativity as well.  It governs the orbits of binary stars, the orbital and escape velocity near self-gravitating masses, the frequency of the fundamental mode of vibration of a self-gravitating mass, and essentially all other processes where self-gravitation determines the structure and dynamics of the system.  If we change the frequency into a velocity,
\begin{equation}\label{eqn:vdyn} 
v_{\rm dyn} = 2\pi f_{\rm dyn}R, 
\end{equation} 
and then we set this to the speed of light, we deduce:
\begin{equation}\label{eqn:blackholecondition}
v_{\rm dyn} = c\qquad \Longrightarrow \qquad R \sim GM/c^2.
\end{equation}
This is, to within a factor of 2 (the Weisskopfian again) the equation for an object whose gravitational escape speed is the speed of light: a black hole.  

\subsection{Radiation: the quadrupole formula}

Radiation of gravitational waves is, to a first approximation, given by the quadrupole formula, which gives the metric $h_{jk}$ of the wave at a distance $r$ from its source in terms of an integral over the source, which is assumed to be described well enough by Newtonian gravity:
\begin{equation}\label{eqn:quadform}
h_{jk} = \left[\mbox{\bf Transverse projection of:}\right] \;\frac{2G}{c^4r}\oderivn{}{t}{2}Q_{jk},
\end{equation}
where the reduced or trace-free quadrupole tensor $Q_{jk}$ (sometimes also called $\ibar_{jk}$) was 
defined in G~Sch\"a\-fer's first lecture at this meeting:
\begin{equation}\label{eqn:defineibar}
Q_{jk}=\int \rho \left(x_j x_k - \frac{1}{3} x^2 \delta_{jk}\right) d^3x.
\end{equation}
The operation called ``transverse projection'' in \Eqref{eqn:quadform} means that only the components of the integral are passed through, and the rest are set to zero.  This enforces the property that gravitational waves act only in the plane transverse to the direction the wave is travelling.

For estimation purposes we shall use a simpler version of this formula which ignores all the indices and makes order-of-magnitude estimates of the integrand in \Eqref{eqn:defineibar}.  Ignoring the indices means that we get {\em upper limits} on the amplitude of the waves, since the projections and the removal of the trace (the term containing the $1/3$ term in \Eqref{eqn:defineibar}) can eliminate components that our estimate will include.  The simpler estimator is:
\begin{equation}\label{eqn:estimate1}
\left|\ddot{Q}_{jk}\right|\leq\oderivn{}{t}{2} \int \rho x^2 d^3x \sim \int \rho v_{\rm dyn}^2 d^3x \sim M v^2,
\end{equation}
where $M$ is the total mass of the source and $v{\rm dyn}$ is given by \Eqref{eqn:vdyn}.  

Of course, this is an upper limit because not all the mass needs to move in such a way
that it gives off gravitational radiation.  Spherical motions, for example, radiate nothing.  One way 
of approximating \Eqref{eqn:quadform}, then, would be to take only the {\em nonspherical} part 
of the kinetic energy $Mv^2/2$ in \Eqref{eqn:estimate1}, which leads to 
\begin{equation}\label{eqn:thorne}
h\sim \frac{4G}{c^4r}K_{\mbox{\small nonspherical}},
\end{equation}
where $ K_{\mbox{\small nonspherical}}$ is the non-spherical part of the system's kinetic 
energy.  This is a good generally-applicable estimate (\cite{thornesnowmass}).  

If all the mass of the system is involved in the motion, and the velocity is determined by self-gravity 
(this excludes radiation from a small 
lump on a spinning neutron star, where only the mass of the eccentricity radiates and the relevant 
velocity is the rotation speed of the star, which could be much smaller than the dynamical speed), 
then we can use the virial theorem to simplify this even further, giving us an {\em upper bound} 
that is usually fairly close to the correct value for sources in the LISA band:
\begin{equation}\label{eqn:hbound}
h \leq \frac{2G^2M^2}{c^4rR} \sim 2\frac{GM}{Rc^2}\frac{GM}{rc^2}.
\end{equation}

This is a very simple formula that was first derived in the context of a scalar approximation to relativistic gravity (\cite{AJP}).  It gives an upper limit on the gravitational wave amplitude in terms of the product of two (dimensionless) Newtonian gravitational potentials: the typical internal potential of the system, $GM/Rc^2$, and the external potential at the observer's location, $GM/rc^2$.  Since the internal potential must be smaller than about 1 (or the system would form a black hole), we see that the gravitational wave amplitude must be smaller than the dimensionless Newtonian potential of the system: waves are a small disturbance in the Newtonian field, not a replacement of it.  

This formula only gives an upper bound on the wave amplitude, but this is not as bad as it might seem.  The real amplitude can fall below this only if the source has some kind of symmetry that does not allow it to radiate fully (such as a nearly-spherical system), or if the frequency is not given by the natural frequency but by a smaller internal frequency, such as the rotational frequency of a spinning star.  But for highly asymmetric source, and especially for the binary systems that are important sources for LISA, this formula is not an upper bound: it is a realistic estimate.

Normally, the frequency of the radiation is twice the natural frequency of the system, essentially because if $v$ depends on time as $\exp{2\pi i f  t}$ in \Eqref{eqn:estimate1}, then the factor of $v^2$ in the integrand has time-dependence $\exp{4\pi i f  t}$. It is not always the case that gravitational waves 
come off at twice the natural dynamical frequency,  but these exceptions need not concern us here.  
Accordingly, we will take 
\begin{equation}\label{eqn:fgw}
f_{\rm gw} = 2f_{\rm dyn} \sim \fracparen{GM}{4\pi R^3}^{1/2}.
\end{equation}
With some interesting values for LISA sources, \Eqref{eqn:hbound}  becomes
\begin{eqnarray}
h&\leq&2.6\E{-22}\fracsqb{M}{2\msolar}^2\fracsqb{R}{2\E8\units m}^{-1}\nonumber\\
&&\qquad\times\fracsqb{r}{10\units kpc}^{-1}\nonumber\\
&&\mbox{\small ($10^3$~s compact binary at galactic centre).}\label{eqn:hbinarygc}\\
h&\leq&2\E{-20}z\fracsqb{M}{2\E6\msolar}^2\fracsqb{R}{6\E9\units m}^{-1}\nonumber\\
&&\mbox{\small (massive bh-binary at redshift $z=1$),}\label{eqn:hmbh}
\end{eqnarray}
where I have assumed a value for the Hubble parameter of $H_0 = 60\units km\;s^{-1}\;Mpc^{-1}$.

\subsection{Energy loss to radiation}

Waves carry off energy, and this is important for some of the systems we will discuss.  One might think that this would be a hard thing to estimate, but this is not really the case.  Relativists argued for decades over whether gravitational waves really did carry energy, because when one looks at the question in the full nonlinear theory of general relativity it becomes a difficult one.  But work in the 1950s and 1960s by H.~Bondi, R.~Penrose, R.~Isaacson, S.~Chandrasekhar, and others put the arguments to rest by showing that general relativity does indeed transmit energy from one place to another via gravitational radiation, and in fact that the formula for the amount of energy is very similar to those in other classical field theories of physics --- electromagnetism and scalar fields, for example.  

In particular, the energy flux carried by a wave is proportional to the square of the time-derivative of the amplitude\footnote{In electromagnetism, for example, the relevant field is the vector potential, and 
its time-derivative is proportional to the electric field in the wave, $\vec{E}$.  Then the rule given 
here would make the flux proportional to $\vec{E}\cdot \vec{E}$, which is the magnitude of the Poynting 
flux {\em for a wave}, up to constants of proportionality.}  $h$:
\[F \propto \dot{h}^2.\]
The constant of proportionality must get the dimensions right, and it can only be made up of pure numbers and the fundamental constants $G$ and $c$.  Remembering that $h$ is dimensionless, the dimensions of $F$ (energy per unit time per unit area) determine the way it depends on $G$ and $c$:
\[F \propto \frac{c^3}{G} \dot{h}^2.\]
The remaining constant is not something that can be deduced by analogy with other theories: it is the only part of this formula that comes from the full tensor theory.  I simply quote it here without proof:
\begin{equation}\label{eqn:flux}
F_{\rm gw} = \frac{1}{32\pi} \frac{c^3}{G} \dot{h}^2 \qquad \mbox{for each polarisation.}
\end{equation}
Because the constant $c^3/G$ has a large value in SI units, this flux can be surprisingly large.  For 
example, for the weakest burst of radiation that the {\em ground-based} detectors anticipate detecting in 
the near future, we use $h=1\E{-22}$ and $f=1\units kHz$.   Then the flux is (allowing for two equally 
strong polarisations)
\[F_{\rm gw} = 3 \units mW\;m^{-2}c\fracsqb{h}{1\E{-22}}^2\fracsqb{f}{1\units kHz}^2,\]
which is twice than the energy flux on Earth from the full Moon.  So for the roughly 1~ms that this source 
is radiating, it is the brightest object in the night sky!  Unfortunately, all that energy goes right 
through the detector, and so the detector's response is woefully small.

Assuming two equally strong polarisations generally (this is still within our factor-of-two uncertainties), 
using $\dot{h} = 2\pi f h$, taking $f = f_{\rm gw} = 2f_{\rm dyn}$ [from \Eqref{eqn:naturalf})], and getting $h$ from \Eqref{eqn:hbound}, we find
\begin{equation}\label{eqn:fluxbound}
F_{\rm gw} \leq \frac{c^5}{4Gr^2}\fracparen{GM}{Rc^2}^5.
\end{equation}
If we again approximate the radiation as being iso\-tro\-pic, we can integrate this over a sphere of radius $r$ to get the total luminosity of the source,
\begin{equation}\label{eqn:luminosity}
L_{\rm gw} \leq \frac{\pi c^5}{G}\fracparen{GM}{Rc^2}^5.
\end{equation}
Notice that this is a very strong function of the internal compactness of the source: a source with $GM/Rc^2\sim 0.2$ (as a neutron star) would radiate $10^{25}$ times the power of one with the compactness of the Sun ($GM/Rc^2\sim 2\E{-6}$)!  The natural luminosity in this equation of 
\[L_{\rm natural}=\frac{c^5}{G}=3.6\E{52}\units W\]
is enormously large, and \Eqref{eqn:luminosity} shows that it is an {\em upper limit} on the luminosity 
of any gravitational wave source. 

For many purposes, the important consideration regarding energy radiated is the time-scale: how long does it take for the gravitational energy loss to manifest itself in a significant way?  The energy is typically lost from the gravitational potential energy of the source $E_{\rm grav} = GM^2/R$ (or from its kinetic energy: to within our factors of two --- Weisskopfians --- these are the same, by the virial theorem).  So the timescale on which observable changes occur is
\begin{eqnarray}
T_{\rm gw} &=& \frac{E_{\rm grav}}{L_{\rm gw}} \geq \frac{R}{\pi c} \fracparen{GM}{Rc^2}^{-3}. \label{eqn:timescale}\\
&\geq &2.4\E6 \fracsqb{M}{2.8\msolar}^{-3}\fracsqb{R}{2\E8 \units m}^{-2} \units yr \nonumber\\
&&\qquad \mbox{(for a 1000~s compact binary)}.\label{eqn:timescalenumbers}
\end{eqnarray}
This gives the timescale as a mulitple of the light-crossing time, $R/c$.  The dynamical timescale $1/f_{\rm dyn}$ might be a more relevant comparison, which leads us to the dimensionless product
\begin{equation}\label{eqn:dynamicaltimes}
T_{\rm gw}f_{\rm gw} \geq \frac{1}{(4\pi^3)^{1/2}}\fracparen{GM}{Rc^2}^{-5/2}.
\end{equation}

\begin{figure*}[!ht]
   \begin{center}
   \leavevmode
 \epsffile[120 300 460 550]{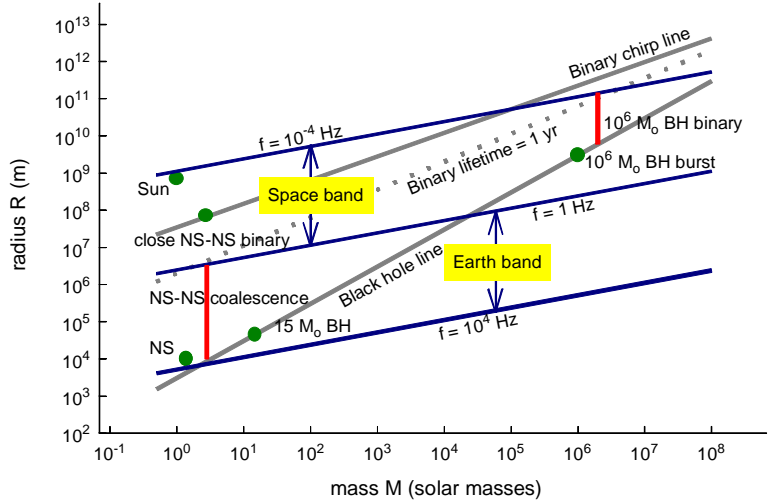}
   \end{center}
\caption{\em This diagram shows the wide range of masses and radii of sources whose natural dynamical frequency is in the band detectable from space or the ground.  The three heavy lines delineate the outer limits of the space band at gravitational wave frequencies of 0.1~mHz,  1~Hz, and 10~kHz.   The ``black hole line'' limits possible systems: there are none below it if general relativity is correct.  The ``chirp line'' shows the upper limit on binary systems whose orbital frequencies change (due to gravitational-wave energy emission) by a measurable amount (30~pHz) in one year: any circular binary of total mass $M$ and orbital separation $R$ that lies below this line will ``chirp'' in a 1-year observation, allowing its distance to be determined.  The curve labelled ``binary lifetime = 1~yr'' is the upper limit on binaries that chirp so strongly that they coalesce during a 1-year observation.  These lines and the indicated sources are discussed more fully in the text.}
\label{fig:sourcedynamics}
\end{figure*}

This has an interesting consequence.  We will see in my second lecture on data analysis that the detectability of a long-duration gravitational wave improves with essentially the number $N$ of observed cycles of the wave, in such a way that the effective amplitude is $h\sqrt{N}$.  Now, if a source can be observed for as long as the gravitational-wave timescale $T_{\rm gw}$, then $N$ is given by \Eqref{eqn:dynamicaltimes}, and we have [using \Eqref{eqn:hbound}]
\begin{equation}\label{eqn:heff}
h_{\rm eff} \sim \fracparen{GM}{Rc^2}^{-1/4} \frac{GM}{rc^2}.
\end{equation}
This has a very weak dependence on the compactness $GM/Rc^2$ of the source.  Somewhat remarkably, then, {\em provided a source can be observed for as long as the gravitational wave timescale, its detectability does not depend strongly on how highly relativistic it is.}  

\subsection{Dynamics in a nutshell}

The most important formulas above can be summarised in a single diagram, which shows a number of relevant lines as a function of the mass $M$ and size $R$ of a the source. \Figref{fig:sourcedynamics}  shows lines of constant frequency $f_{\rm gw} = 2f_{\rm dyn}$  in the mass-radius plane for 3 important frequencies: $10^{-4}$~Hz, the lowest frequency accessible to LISA; 1~Hz, roughly the boundary between what can be detected from the ground and from space; and $10^4$~Hz, the upper limit to what can in practice be observed from the ground.  The upper part of the diagram is therefore the space-accessible region; the lower part, the domain of ground-based detectors. 

In the diagram we place a number of interesting possible gravitational-wave sources.  At the low-mass end, the natural vibrations of a typical neutron star and stellar-mass black hole radiate in the ground-based band; these should be excited when the objects are formed.  The Sun lies in the space band, and indeed its natural vibrations could be detected by a space detector, through the near-zone Newtonian gravitational oscillations they produce rather than through their gravitational waves.  Binaries in this mass range are discussed below.  At the high-mass end, a $10^6\msolar$ black hole would radiate in the space band.  These vibrations could be excited by the formation of the hole or by a neutron star falling into such a hole.

There are other useful lines in this diagram, as described in the next sections.  

\subsubsection{The black-hole line}

The most important is the black-hole line, drawn for 
\begin{equation}\label{eqn:bhline}
R=\frac{2GM}{c^2}=3\E9\fracsqb{M}{10^6\msolar}\units m.
\end{equation}
(Compare this with \Eqref{eqn:vdyn}).  The region of the diagram below this line does not contain any physically realisable systems: a system forms a black hole when it reaches this line from above.  The space-accessible frequency region contains black holes above about $10^4\msolar$ up to $10^8\msolar$, which means that space detectors can in principle confirm the present astrophysical consensus that most galaxies contain one or more giant black holes.  Conversely,  ground-based detectors cannot see massive black holes, being limited to observing the kind we expect to form from normal massive stars.  
	
\subsubsection{Binary lifetime line}

Two other lines in the diagram refer to the chirping of a binary system, as discussed above. The line called ``binary lifetime = 1~yr'' is the line along which the characteristic timescale for the frequency to change, as inferred from \Eqref{eqn:timescale}, is one year.  Binary systems below this line are systems which can be followed right to coalescence during a reasonable observation period, and whose detectability is therefore not strongly dependent on how compact they are when they are first observed.  From \Eqref{eqn:timescalenumbers}, we see that this is a line on which $R^4/M^3$ is constant. Notice that {\em all} solar-mass binary systems observable from the ground will coalesce within a year.  A typical coalescing neutron-star binary is illustrated in the diagram.  A compact binary that is observed 
from the time it reaches about 1~Hz will coalesce within a year.  At present, no detectors are planned 
which can operate well at this frequency.  If one were available, it could give advance warning to existing detectors about coalescence events.  From space, we can expect only binaries of massive black holes, above $M \sim10^6\msolar$, to coalesce during an observation, as shown.  

\subsubsection{Binary chirp line}

Just as important, but less dramatic, is just seeing a binary system ``chirp'', i.e. change its orbital frequency.  Here the criterion is not that its coalescence time-scale be the observation time, but rather that its frequency should change by an observable amount during the same one-year observation 
$T_{\rm obs}$.  This means that its frequency change need only be as large as the frequency resolution 
of a 1-year observation,  $\Delta f_{\rm gw} = 1/T_{\rm obs} =3\E{-8}\units Hz$.   If we take the frequency 
change to be the same $\Delta f$, and assume that this occurs because of gravitational radiation, then 
we have 
\[\Delta f_{\rm gw} = \frac{f_{\rm gw}}{T_{\rm gw}}T_{\rm obs}.\]  
The formulas above can be used to show that the resulting ``chirp line'' is a line of constant $R^{11}/M^7$.  For a separation and mass appropriate to a compact binary with a 1000~s period, we 
have
\begin{equation}\label{eqn:chirpintimeT}
\fracsqb{R}{2\E8\units m}^{11}\fracsqb{M}{2.8\msolar}^{-7} = \fracsqb{T_{\rm obs}}{3.7\units yr}^4.
\end{equation}

The diagram shows the chirp line appropriate to a 1-year observation, essentially the same as \Eqref{eqn:chirpintimeT}..  It shows that chirping without coalescence is important for space-based detectors; ground-based detectors will be able to follow any chirping system right to coalescence. It is 
clear that a good fraction of binaries that LISA will observe will chirp during an observation.  As we 
show below, this allows LISA to determine the distance to the binary.  
A space detector should also detect chirping in binaries consisting of massive black holes.  The resulting distances will be particularly interesting for LISA, as we describe below.  

\subsubsection{Distance to a chirping binary}

The key to determining the distance is to show that there are enough observables in the signal from 
a binary system to make the measurement.  The mass and radius of the system, which are convenient 
axes for the diagram in \Figref{fig:sourcedynamics}, are not directly observable.  What we can determine 
from the response of the detector are the frequency $f_{\rm gw}$, rate of change of frequency 
$\dot{f}_{\rm gw} = f_{\rm gw}/T_{\rm gw}$ (if the system chirps) and amplitude $h$ of the signal.  
Equations~\ref{eqn:fgw}, \ref{eqn:timescale}, and \ref{eqn:hbound} (taken as an equality, which 
is okay for a binary system) together allow us to eliminate all the unknowns and solve for 
the distance $r$ to the binary.  This gives the remarkably simple formula
\begin{equation}\label{eqn:distancetobinary}
r = \frac{c}{2\pi^2}\dot{f}_{\rm gw}f_{\rm gw}^{-3}h^{-1}.
\end{equation}

This equation, first derived in a somewhat different form by \cite{schutz}, is actually more robust than 
our simple derivation might suggest.  We have used a single mass $M$ to characterise the system, 
but of course a binary has two masses.  Which combination of them is appropriate here?  More 
importantly, can we really eliminate the mass at all from these equations: maybe we have to 
eliminate two masses, and we don't have enough equations.  

The answer is that there is only one mass that matters, which is the combination 
\begin{equation}\label{eqn:chirpmass}
\mptr = \mu^{3/5}M^{2/5}, 
\end{equation}
where $\mu$ is the reduced mass of the binary and $M$ its total mass.  Our analysis here was not detailed enough to distinguish these two masses, but if we had done so then we would have found that this is the way the masses of the individual stars enter the radiation timescale equation, Eqref{eqn:timescale}, if we eliminate the unknown radius $R$ in favour of the measureable frequency $f_{\rm dyn}$.  This gives a relation between the measured timescale and frequency and the mass of the system, which then can be used to determine the chirp mass $\mptr$.  

What is remarkable about binaries is that the same chirp mass also enters the equation for the amplitude of the radiation, again obtained by eliminating $R$ in favour of $f_{\rm dyn}$ in \Eqref{eqn:hbound}.  Then one can go through the same procedure, only using $\mptr$ in place of $M$ 
in all our equations, and arriving finally at the distance $r$ to the binary given by \Eqref{eqn:distancetobinary}, with $M$ replaced by $\mptr$.  

For cosmological sources, this distance turns out to be what cosmologists call the luminosity distance.  The ability to treat chirping binaries as standard candles is one of the most interesting aspects of gravitational wave observations.  It opens the possibility of using observations of chirping systems 
to measure the Hubble parameter $H_0$ (\cite{schutz}) and even the deceleration parameter of the universe $q_0$.

\section{LOW-FREQUENCY SOURCES DETECTABLE BY LISA}\label{sec:low-f}

\subsection{LISA's capabilities}

Sources of gravitational waves that emit in the low-frequency regime accessible from space are mainly either stellar-mass binary systems with relatively large separations and therefore weak gravitational fields (far from the black-hole line in \Figref{fig:sourcedynamics}), or massive systems that are highly relativistic and therefore almost inevitably contain black holes.  The exceptions are the Sun (which influences a detector through its time-dependent near-zone Newtonian field) and a random (stochastic) background of gravitational waves that merely appears like an extra noise in the detector. 

\begin{figure*}[!ht]
   \begin{center}
   \leavevmode
\epsffile[130 300 490 560]{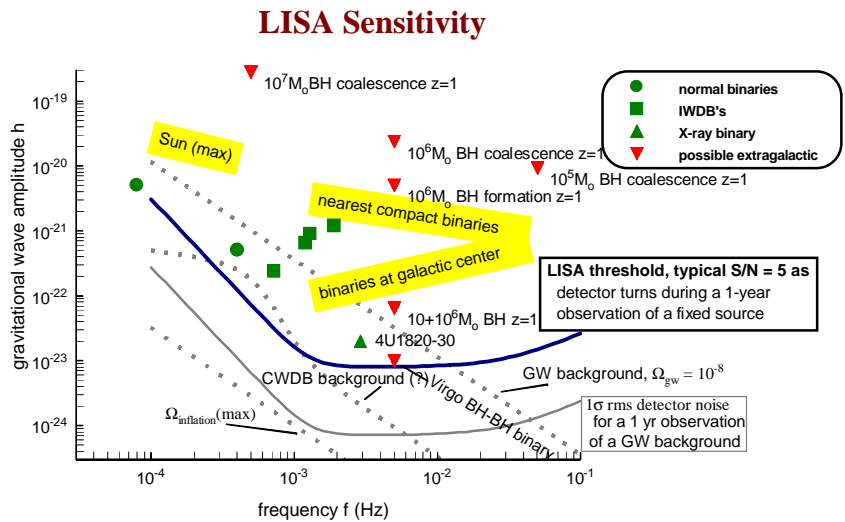}
   \end{center}
\caption{\small\em Strength of various sources and the sensitivity
  curve of LISA: plot of the intrinsic amplitude of likely 
  \gw s against their frequency.
  Most LISA sources will be approximately monochromatic.
  The bold curve is the 1-year sensitivity curve,
  the amplitude that could be detected by a single (2-arm) LISA interferometer 
  in a 1-year observation with confidence, {\em i.e.} with  
  a signal-to-noise ratio of 5, allowing for the rotation of LISA during an 
  observation. Below it, drawn faintly, is the actual r.m.s.\ noise level
  for a 1-year integration.
  The \gw\  amplitude $h$ is shown for different types of
  periodic and quasiperiodic sources.  
  The strongest sources in the diagram are coalescences of binaries of 
  massive black holes at cosmological distances.  They have been placed 
  in the diagram at their coalescence frequency, at 
  an amplitude that correctly shows their signal-to-noise ratio in relation to the LISA 
  sensitivity curve, for a distance $z=1$ with $H_0=75\units  km\;
  s^{-1}Mpc{-1}$.    
  The expected signals from 
  some known binaries are indicated by circles and boxes, as 
  identified in the text. The nearest neutron-star 
  and white-dwarf binaries at any frequency should 
  lie in the band labelled ``nearest compact binaries''; the band below that 
  shows the amplitudes expected from ``typical'' white-dwarf 
  binaries near the galactic centre. 
  A possible cosmological background left  from 
  the Big Bang is shown here at  
  an energy density per decade of frequency today that 
  is $10^{-8}$ of the total needed to close the Universe, again for 
  $H_0=75\units  km\;  s^{-1}Mpc{-1}$. A possible upper limit to 
  that generated by inflation is 
  also shown.  If there is a 
  confusion limit due to galactic binaries, as discussed in the text, 
  then it might appear as shown. 
  The band labelled ``Sun (max)'' is where solar g-modes might produce strong 
  near-zone (Newtonian) gravitational perturbations observable by LISA.
 (Figure reproduced from (\protect\cite{LISA1996}).)
  }
\label{fig:lisa}
 \end{figure*}

All of these sources tend to be long-lived.  Even a black-hole coalescence has a natural timescale of several months, and nonrelativistic binaries in the low-frequency window live for thousands or millions of years.  Therefore the appropriate way to display the strength of a source compared to the sensitivity of a detector is to assume, say, a 1~yr observation time, and show the intrinsic amplitude of the source against the noise of the detector after integrating for a year (see the data-analysis section below).  We draw such a diagram for the LISA detector in \Figref{fig:lisa}.  Shorter-duration events are shown at an amplitude that correctly represents their signal-to-noise ratio if they are extracted by optimal pattern-matching.  Most long-lived sources have a fixed frequency; those that do not are shown at the highest frequency they reach.

In this diagram we show the LISA noise curve in two ways: the lower, lightly drawn curve is the actual noise in a 1~yr integration.  The upper, bolder curve is a fairer representation of LISA's sensitivity.  It is set at a threshold of a $S/N$ ratio of 5, taking into account antenna-pattern effects as LISA rotates during its orbit of the Sun: any source above this threshold will be detectable with confidence over almost the entire sky.  Sources below this line will be detectable if there is independent information about them (such as their exact frequency) that will allow one to lower the confidence threshold.  And when we discuss the cosmological stochastic background, the lightly-drawn $1\sigma$ noise curve is the appropriate comparator.

\subsection{Signal reconstruction}

\Figref{fig:lisa} shows LISA's sensitivity in terms of the amplitude $h$ of gravitational waves, but 
what does $h$ mean?  What LISA directly records is a strain difference between any two arms, say 
$(\delta \ell/\ell)_1-(\delta \ell/\ell)_2$.  This is related to $h$ by a set of projections: first, the projection
of the plane of the gravitational wave onto the plane of the detector (this depends on the position of 
the source in the sky) and second the projection of the intrinsic polarisation of the wave onto the polarisation of the particular detector.  This latter depends on the orientation of the source relative 
to the line of sight to it: a binary system seen down its orbital axis radiates circularly polarised waves, 
while a binary seen in its orbital plane radiates plane-polarised waves.  

For the purpose of inferring the distance to a chirping binary, as described above, we must know the 
intrinsic amplitude of the wave, as radiated by its source.  This means having certain information about 
the projections: we must know the direction to the source and the angle of inclination and orientation of 
the plane of its orbit.  To get this information, LISA must use more than just a single number $h$ 
characterising its response in some average way, even though that is all we have shown in 
\Figref{fig:lisa}.

LISA makes use of the phase and amplitude modulation of signals to return directional and polarisation information.  For weak sources, phase modulation provides no resolution below about 1 mHz, where the gravitational wavelength becomes comparable to the orbital radius of 1~AU. Accuracy should be at the level of a few tens of degrees or better above this (\cite{peterseim}).  Binaries in our galaxy could be located to within tens of arc minutes, depending on signal-to-noise ratios.  A recent detailed study by Cutler and Vecchio (\cite{vecchio}) shows that LISA will attain angular accuracies of order one degree on the most interesting massive-black-hole binary sources, but it may not be able to reach much below that.  Amplitude modulation can always be used to determine not only the polarisation of the signal.  

The second LISA interferometer provides independent information about the signal.  If the signal is strong, then even if there is common noise between the two interferometers, the information will not be lost, and we can treat the two interferometers as genuinely independent.  In this case, there is direct polarisation information in the two different signals, and this can be combined with the amplitude modulation to improve the polarisation sensitivity and the directional information (\cite{jennrich}). 

I will discuss the way this is done in my second lecture at this school, on data analysis for low-frequency gravitational waves.

\subsection{Binary systems in the Galaxy}

One of LISA's main targets will be galactic binary systems, particularly those containing neutron stars and/or white dwarfs.  This subject has been reviewed recently by Verbunt (\cite{verbunt}).  Although all 3 known Hulse-Taylor-type pulsar systems emit orbital gravitational radiation at frequencies somewhat too low for LISA, LISA has much greater range than radio pulsar surveys. The statistical analyses of pulsar binary observations (\cite{lorimer}) suggest that there should be of order 100 neutron-star binaries in the Galaxy within the LISA frequency range, and LISA would be able to see them all.  If rather larger estimates based on evolution calculations (\cite{lipunov}) are right, then there may be thousands, and one or two may even lie below the ``chirp line'' in \Figref{fig:sourcedynamics}.  LISA will be able to determine their distance directly from the chirp information.  \Eqref{eqn:hbinarygc} shows the amplitude 
expected from a neutron-star binary at the galactic centre.  It falls in the indicated band in the diagram.

With so many neutron-star binaries visible, there should accordingly also be tens or hundreds of neutron-star/black-hole binaries and perhaps a similar number of black-hole/black-hole systems, which will be stronger and therefore even more easily detectable.  In fact, it is likely that there will be of order one black-hole/black-hole binary at a high-enough frequency to be visible from the Virgo cluster, which is shown in the diagram.

There are other binaries that ought to be even more plentiful. Some known X-ray binaries and cataclysmic variables are in the range of LISA; in fact. if they were not detected by LISA, it would be disastrous for general relativity.  Many of the Thorne-Zytkow stars in which ordinary pulsars may be spun up to millisecond periods could also radiate in the LISA band as the neutron star orbits the core of a giant star inside the envelope of the giant (\cite{postnov}).

Importantly, there should be a large number of white-dwarf binaries, which are very difficult to detect by optical observations. Our present observational limits on their population are weak, and it is possible (or even likely) that they will be so plentiful that they will provide a confusion-limited background at low frequencies (\cite{Hils1990,bender}).

The information we would get from observing these systems would have some relevance to fundamental physics, but not as much as the black-hole systems we will discuss below.  One interesting possibility is the detection of a scalar component to the gravitational field, which might arise in a superstring-unified field theory in which gravitation is just one aspect of the interaction of elementary particles.  Scalar components are radiated away when black holes form, so they have no effects on two-black-hole systems.  Similarly, scalar radiation is suppressed on grounds of symmetry if the two stars in a binary are of similar mass and compactness, as in the Hulse-Taylor pulsar system.  However, a binary consisting of a neutron star and a lighter white dwarf might well show the effects of scalar radiation, both in the detailed dynamics of the evolution of its orbit (due to modifications in \Eqref{eqn:luminosity}) and in the directly detected polarisation pattern of the radiation.  The system would have to be relatively nearby to give LISA a good signal-to-noise ratio, but this is not highly improbable.

LISA observations of binaries would provide a rich harvest of astrophysical returns. One of the most interesting pieces of information would be the polarisation of the signal. This will tell us the three-dimensional inclination of the orbital plane. For a known binary, whose mass function is known from spectroscopic observations. and whose primary mass is estimated from models, knowing the inclination determines the mass of the secondary. Then the intrinsic amplitude of the gravitational waves from the system will determine the distance to the binary.  This extra information will be crucial for modelling such systems.  

\subsection{Massive black holes}

The model that active galactic nuclei contain supermassive black holes has gained wide acceptance among astronomers in the last decade. These holes may have masses up to $10^9\msolar$  or more.  But active nuclei are rare, and most galaxies may have seen only modest amounts of activity in their past.  However, there is growing evidence that ordinary galaxies and perhaps even dwarf galaxies commonly contain more modest black holes in the mass range $10^5$--$10^7\; \msolar$ (which we shall refer to as simply massive black holes, in contrast with supermassive black holes).  As \Figref{fig:sourcedynamics} shows, this is the mass range that a space-based detector would be sensitive to.  The supermassive holes radiate at too low a frequency, but the massive black-hole range can radiate at LISA frequencies in at least 3 ways: as binary systems that coalesce, as large black holes that accrete smaller-mass compact stars and holes, or in the process of formation of the holes themselves.  In a later lecture, G Sch\"afer will review the evidence for these holes.  There is also a recent published discussion of the gravitational radiation from them in the LISA context (\cite{rees}).  

As an example, \Eqref{eqn:hmbh} shows what amplitude might be expected from a massive black 
hole binary at the distance corresponding to a redshift of 1.  This rough amplitude does not take 
into account any cosmological effects, such as the redshifting of the frequency.  Nor does it take 
into account signal enhancement by matched filering (see below).  Therefore it is only indicative 
of the level at which we might see such binaries.

\subsubsection{Binary merger of massive black holes}

A binary merger of two massive black holes is the strongest source we anticipate for LISA, and one of the most interesting.  A merger at a redshift of 1 of holes with a mass $10^6\msolar$ would have an amplitude signal-to-noise ratio of perhaps $10^4$, depending on how much background there was from galactic binaries (see below).  This is the signal-to-noise ratio that could be obtained by matched 
filtering of the data stream, looking for a signal with the expected waveform, as described in my 
second lecture at this meeting.  The position in \Figref{fig:lisa} where this event is shown is not 
meant to be exact.  The actual waveform scans through a range of frequencies, and the 
signal builds up as we follow this.  So the points plotted for such events in the figures only 
show the final detection signal-to-noise, and are plotted at the coalescence frequency.

Such signals are so strong that LISA will be able to locate their sources with errors less than 1 degree (\cite{vecchio}).  Combined with the apparent amplitude of the signal and its polarisation as measured 
by the amplitude modulation and the independent data in the two LISA detectors, this will give the true intrinsic amplitude to accuracies perhaps as good as 1\%.  

Such events are not necessarily rare.  If most galaxies have moderate-mass black holes, then maybe some have more than one, either because they form in binaries like stars typically do, or because they are brought together by the merger of their original host galaxies, an event that was probably common early in the Universe.  The statistics are difficult to estimate (\cite{rees,vecchioLISA}), but it seems likely to me that the event rate will be either one every few years or several per year.  Only observations will tell.

At a redshift $z=1$, the angular accuracy of LISA corresponds to an error box containing only a few rich clusters of galaxies, and perhaps fewer active/\-interacting/\-peculiar galaxies that would be candidate hosts for this event.    The combination of the redshift obtained from even a tentative optical identification and the distance to the source provided by LISA itself from the chirp-rate of this signal would allow a determination not only of the Hubble constant --- which ought by then to be known by other means --- but even more importantly of the deceleration parameter $q_0$ of the Universe.  In principle it could get both of them to the unprecedented accuracy of perhaps 1\%.  From this one can infer the mass density of the Universe or decide whether there must be a cosmological constant.  This is certainly one of the most fundamental observations that LISA could make.

Moreover, with such good signal-to-noise ratios, LISA could compare the details of the merger phase of two black holes with the results of numerical simulations of black-hole coalescence.  These would test both the numerical codes and the correctness of general relativity as a descriptor of strong-field gravity.  With good signals, one could test such theorems as the Hawking area theorem, which says that the total area of the horizons of a system of black holes cannot decrease.  This is the classical manifestation of the fact that quantum effects give black holes an entropy.  

\subsubsection{Compact stars falling into massive black holes}

Massive black holes in galactic centres should also occasionally swallow up stars. While main-sequence stars and giants are so large that they will be torn apart by tidal forces before they reach the horizon, neutron stars and stellar-mass black holes will remain essentially point particles that follow very complex orbits until they finally fall into the hole.  These are not easy to model exactly, but with approximate matched filters that follow portions of the orbit, it should be possible to see these event at redshifts of 1 or so. They should be more plentiful than black-hole mergers, and event rates of several per year seem likely (\cite{sigurdsson}).

One can calculate an approximate signal strength by the formulas we derived earlier, but one should 
in this case be careful to keep the mass of the radiating star separate from that of the hole:
\[h\leq2\frac{GM_{\rm bh}}{c^2R}\frac{GM_{\rm star}}{c^2r}.\]

These events are particularly interesting because they can also test strong-field gravity near black holes.  The orbit of a compact mass near a large black hole is a good approximation to a geodesic, and the geodesics near a black hole are good probes of its geometry.  In order even to detect these events, the signals must be fit to a model of the orbit that includes all the physical influences of the black hole, such as the dragging effect of its spin.  Gravitational spin-orbit and spin-spin coupling should be observable.  Such observations can determine if the massive hole is well-described by the Kerr metric, which in general relativity is the unique solution for a spinning vacuum black hole.  These fundamental tests will provide the most stringent examination of classical general relativity imaginable.  

\subsubsection{Gravitational collapse to massive black holes}

If massive black holes form in one go, from the collapse of a cloud of gas or small stars, and if that collapse is highly non-symmetrical, then there could be significant events of limitied duration in LISA data.  At present, the astrophysical evidence leads one to think that such event occur but that their asymmetry may be very small (\cite{reesUCP}).  But if there are many such events per galaxy, for example if galaxies are made of mergers of smaller clouds of gas of $10^*\msolar$ or so, and if each such cloud has its own black hole, then there may be an observable fraction of such events.  I have plotted them on the diagram, since they represent such a strong and unmistakable signal, even at cosmological distance.

\subsection{Stochastic Background: Our Earliest View of the Big Bang}

Just as the Big Bang left us the cosmic microwave background radiation, so too is it likely to have left a background of gravitational radiation.  Because gravitational waves interact so weakly with other matter, this radiation is genuinely primordial: apart from a cosmological redshift, it is unchanged since it was produced.  This is an important difference from the microwave background, which was thermalized and strongly coupled to matter until the epoch of recombination.  While the microwave background comes to us from about $10^5$ years after the Big Bang, any gravitational wave background will come directly from a much earlier time, possibly only $10^{-25}$~s after the Big Bang.  

The radiation is stochastic, consisting of many individual components superimposed on one another in a random way.  It can be adequately characterised by its energy density as a function of frequency.  The conventional way of doing this is: 
\begin{itemize}
\item First, define  $\rho_{\rm gw}(f)$ to be the contribution to the cosmological energy density from gravitational waves up to frequency $f$. 
\item Second, define an energy density per unit logarithmic frequency ({\em i.e.} between any frequency $f$ and $e$ times that frequency) by taking the logarithmic derivative of this density.
\item Third, normalise this to the energy density that is required to close the Universe, $\rho_c = 2\times 10^{-43}\units J\; m^{-3}$ (for a Hubble constant of $100\units km \;s^{-1}\; Mpc^{-1}$).
\end{itemize}
This gives
\begin{equation}\label{eqn:omegastochastic}
\Omega_{\rm gw}=\frac{f}{\rho_c}\oderiv{\rho_{\rm gw}}{f}.
\end{equation}

This definition is a natural one for radiation that is produced by physical processes that have no natural length-scale, so that $\rho_{\rm gw}$ depends only on a power of $f$.  For such scale-free radiation, the energy density $\Omega_{\rm gw}$ will depend on the same power of the frequency.  Most models of Inflation  produce approximately a constant-$\Omega_{\rm gw}$ spectrum, which is called a {\em Harrison-Zel'dovich} spectrum, but some models produce much more complicated spectra.    Recent models based on string theories (\cite{BV1,BV2,Buonanno1996}) suggest a spectrum for which $\Omega_{\rm gw}(f)\propto f^{3}$ over a certain spectral range that depends on the details of the physics generating the waves.  In this case, the {\em amplitude} $h$ rather than the logarithmic energy density is independent of frequency.  Such a spectrum could  favour detections by ground-based detectors if they are fortunate enough to have the rising part of the spectrum in their sensitivity range, but it could equally well favour LISA.

Inflation produces gravitational radiation by parametric amplification of quantum fluctuations that are assumed to exist at the Big Bang.  As was first shown by Grishchuk (\cite{Grishchuk}) and Starobinsky (\cite{Starobinsky}),   while inflation --- if it occurred --- was redshifting any thermal background away, it was simultaneously amplifying a much lower-frequency background due to quantum fluctuations.   

COBE observations of the microwave background fluctuations can be adapted to provide upper limits on the amount of radiation that could have been produced by parametric amplification of quantum fluctuations by inflation. This is shown in the figure, and it cuts close to but below the LISA noise curve.  However, as mentioned above, there is very likely to be a background  due to white-dwarf binaries at this frequency that will overwhelm this cosmological one, in which case no detector would be able to see a cosmological background this weak.  Grishchuk has more optimistically estimated that COBE's observed spectrum, when extrapolated to LISA's band, could be very detectable, even as large as $10^{-8}$ (\cite{grishchuk}).

An alternative to inflation is the possibility that topological defects were produced by the spontaneous symmetry breaking that produced the low-energy physics (separation of forces, masses of particles, etc) that we observe today.  If the symmetry breaking followed the rules we understand for gauge theories like the electroweak interactions, then it is possible (depending on the symmetry group being broken) that this process has left behind some remnants, or defects, that contain a memory of the unbroken state.  These defects may be regions of high mass concentration, and they may give off gravitational waves.  If the defects were common enough, such waves would form a random background today.

At a frequency of $10^{-3}$~Hz today, in the LISA band, the radiation would have been emitted at the time the Universe was going through the electroweak phase transition, when the typical energy was about 100~GeV (\cite{Allen1996}).  Theorists do not believe that this transition was strong enough to produce radiation by itself, but other physics at the time might have led to defects like cosmic strings, domain walls, or textures (\cite{Allen1996}).  Currently, it is felt that such defects cannot explain the structure seen by COBE in the microwave background, and that inflation is the best model for this.  This does not, however, exclude defects as a source of cosmic gravitational radiation at a weak level that would still be detectable by LISA, or indeed by the ground-based detectors.

A background of gravitational waves appears in a single detector, like LISA, as simply another source of noise.  It is not possible to gain significantly in sensitivity by correlating the two ``independent'' LISA interferometers, because they possess a common arm and therefore a common source of instrumental noise.  Provided it is above other noise sources, and provided we have confidence in identifying or limiting other noise sources, then LISA could make deep searches, particularly in the frequency range above 1~mHz.

The impact on theoretical physics of the discovery of a cosmic background of gravitational waves would be enormous.  It would be likely to confirm the existence of inflation or of some topological defect.  The observed spectrum in turn would give definite parameters (energies, symmetries) that would guide the development of models for high-energy physics.  Probably no observation by LISA could have a greater impact on fundamental physics.

\section{CONCLUSIONS}

LISA or another similar space-based gravitational wave detector will open the low-frequency gravitational wave window.  It can explore a variety of systems, from binary systems of compact stars to black holes to the Big Bang itself.  The observations will have profound consequences for fundamental physics, confirming (or otherwise!) the Einstein field equations, discovering or constraining scalar gravitational fields, and possibly providing insight into the earliest moments of the Universe.  LISA is potentially one of the most rewarding space missions for basic physics that has ever been proposed.

\end{document}